\documentclass[pre,reprint,twocolumn,amsmath,10pt,amssymb,aps,superscriptaddress,showkeys,showpacs]{revtex4-1}\input epsf
\usepackage{bm}
\usepackage{soul}
\usepackage{graphicx}
\usepackage{makeidx}
\usepackage{chemarr}
\usepackage{multirow}
 \usepackage{color}
 \usepackage[normalem]{ulem}
\definecolor{darkgreen}{rgb}{0,0.6,0}
 \definecolor{orange}{rgb}{0.99,0.257,0}
\newcommand{\be}{\begin{equation}}
\newcommand{\ee}{\end{equation}}
\newcommand{\ba}{\begin{eqnarray}}
\newcommand{\ea}{\end{eqnarray}}

\def\epsilon{\varepsilon}

\def\i{\indent}

\def\beqr{\begin{eqnarray}}
\def\eqnr{\end{eqnarray}}
\def\beq{\begin{equation}}
\def\bc{\begin{center}}
\def\ec{\end{center}}
\def\eqn{\end{equation}}

\usepackage{amsmath}

\usepackage{graphicx}
\usepackage{dcolumn}
\usepackage{bm}

\begin{document}

\title{Percolation transitions in the binary mixture of active Brownian particles with different softness}
\author{Monika Sanoria}
\email{msanoria@ucmerced.edu}
\affiliation{Department of Physics, Indian Institute of Technology Bombay, Powai, Mumbai, 400076, India.}
\affiliation{Center for Cellular and Biomolecular Machines, University of California Merced, CA, 95343, USA.}
\author{Raghunath Chelakkot}
\email{raghu@phy.iitb.ac.in}
\author{Amitabha Nandi}
\email{amitabha@phy.iitb.ac.in}
\affiliation{Department of Physics, Indian Institute of Technology Bombay, Powai, Mumbai, 400076, India.}

\begin{abstract}
Homogeneous active Brownian particle (ABP) systems with purely repulsive interactions are considered to have simple phase behavior, but various physical attributes of active entities can lead to variation in the collective dynamics. Recent studies have shown that even homogeneous ABPs exhibit complex behavior due to an interplay between particle softness and motility. However, the heterogeneity in composition of ABPs has not been explored yet. In this paper, we study the structural properties of a binary mixture of ABPs with different particle softness, by varying the relative softness and composition. We found that upon varying the motility parameter, the system underwent a motility-induced phase separation (MIPS) followed by a percolation transition similar to the homogeneous systems; however, there is complex structure formation within the dense phase of the MIPS which depends on the relative softness of the binary system. Furthermore, the presence of a non-linear scaling for different compositions of heterogeneous ABPs suggests that there is a complex relationship between the composition and the structural properties. Our study demonstrates that the composition heterogeneity of ABPs can also lead to complex phase behavior.

\end{abstract}

\maketitle

\section{Introduction}
\label{intro}
The emergence of collective behaviour in active matter systems has been a burgeoning area of research for the past several years. Such emergent non-equilibrium behaviour has been observed in a large class of biological and synthetic systems for a wide range of length scales.~\cite{Ramaswamy2010,vicsek2012collective,Marchetti2013,cavagna2010scale,BECCO2008scale,zhang2010collective,beer2020phase,tan2020topological,JULICHER2007,schaller2010polar,sumino2012large,rubenstein2014programmable, KokotPNAS2017, cohenPRL2014,gompper2020, bricard2013emergence, thutupalli2011swarming,Buttinoni2013}. Since many dynamical features of such large-scale behaviour are qualitatively similar across a wide range of systems, simplified particle-based models have been used to study such phenomena. Although such models are minimal and highly simplified in microscopic details compared to the actual systems, they are successful in providing insight into the underlying features that give rise to observed collective behaviour.

  One such particle-based model that has been widely used to explore the density ordering in active matter is active Brownian particles (ABP). In the simplest version of the ABP, the self-propelled particles only interact via short-range repulsive forces, and their propulsion direction evolves diffusively~\cite{Fily2012,Redner2013,catesReview2015,stenhammar2014phase,digregorio2018full}. It is well known that a collection of such ABPs phase-separates into dense and dilute phases at sufficiently large motilities and densities, known as motility-induced phase separation (MIPS). This type of phase separation occurs even in the absence of any attractive interaction between the particles and is purely driven 
  by particle motility. ABP model has been extensively studied for mono-disperse systems, and a detailed bulk phase diagram as a function of density and particle motility has been generated~\cite{Redner2013,Fily2012,digregorio2018full,Cates2015,stenhammar2014phase,speck2015dynamical}. The phase behaviour of these systems are considered to be relatively simple compared to other numerical models, which involve complex alignment interactions ~\cite{kyriakopoulos2019clustering}. However, many of the structural and dynamical properties, especially in the dense phase and under various environmental heterogeneities, are still being revealed~\cite{caprini2020spontaneous, caporusso2020motility, shi2020self, Das2020, Das2020_pre}, thus making such phase-separating systems continues to be an exciting topic for research.

Although ABPs are generally believed to show relatively simple phase properties, recent studies on soft ABPs have shown a more complex phase behaviour~\cite{sanoriaPRE2021, sanoria2022Percolation, de2022collective, martin2022dynamical}. At large motilities, the compact and structurally ordered dense clusters of ABPs are destabilized to form space-filling, porous clusters. The formation of such motility-induced porous clusters has also been reported in ABPs in the presence of  an external potential~\cite{mukhopadhyay2023active}. It has also been shown that the critical exponents associated with this structural transition are in very good agreement with the standard percolation transition in 2D~\cite{sanoria2022Percolation}. Thus, at the high motility, soft interaction limit, the compact, dense clusters are destabilized through a percolation transition, and they are replaced by a spanning cluster. This transition can be observed even at low motility if the interparticle interaction is made softer. This \emph{soft} limit of the active particles is relevant in the context of biological systems since cells are typically deformable. However, all the studies which are described above are conducted in a monodisperse system of ABPs, where the activity, interaction strength and range of interactions are the same for all the particles. While these studies reveal interesting collective properties relevant in many naturally occurring systems, it is imperative to consider the diversity in compositions in mechanical properties of the particles to make the models more useful in studying a wide variety of biological systems. The mechanical properties of cells are, in general, not uniform across tissues. For example, as seen in case of cervical and breast cancer, cells have distribution of rigidity, which plays an important role in tumor progression~\cite{fuhs2022rigid}. The change in the rigidity of the cells significantly impacts their ability to migrate through dense domains ~\cite{seltmann2013keratins}. While the systems with diversity in mechanical properties are yet to be explored in detail, heterogeneity in other microscopic features is studied in some detail. For example, previous studies on systems with heterogeneity in activity~\cite{dolai2018phase, kolb2020active}, chirality~\cite{huang2020dynamical}, size polydispersivity~\cite{yang2014aggregation, fily2014freezing} have revealed many novel collective features.

In this study, we explore the effect of heterogeneity in interaction softness in the phase behavior of active Brownian particles (ABPs). We consider a binary mixture of particles with different interaction softness and study the large-scale properties as a function of the composition and particle motility. We specifically examine the overall phase properties of the binary mixture, as well as the spatial arrangement of the types of particles within various phases. 


\section{Model}

We consider a 2D system containing both $N_h$ \textit{hard} and $N_s$ \textit{soft} particles uniformly distributed in a periodic box of length $L$. Each {\it i}th particle is characterized by its position ($r_i$), stiffness ($k_i$) and orientation ($\theta_i$).
The dynamical equations of motion are given by the overdamped Langevin equations as 
\begin{align}
    \dot{\mathbf r}_i &= \mu~  \mathbf{F}(\mathbf r_{i}) + {v_p}~\hat{\mathbf{n}}_i \\ \dot{\theta}_i &= {\xi}_i^R
\end{align}
with $v_p$ is the self-propulsion speed, $\mu$ is the mobility and $\hat{\mathbf{n}}_i = (\cos{\theta_i}, \sin{\theta_i})$ is the instantaneous direction of the $i^{th}$ particle. The Gaussian white noise $\xi_i^R$ has mean zero, and follows the relation $\left<\xi^R_i~\xi^R_j\right> = 2~D_r \delta_{ij}~\delta(t-t')$. 
The force acting on the $i^{th}$ particle is given as
\begin{equation}
    \label{Vex}
 \mathbf{F}(\mathbf r_{i}) =
\begin{cases}
-\sum_{j \neq i}^N k_{eff}(ij)~(\sigma - {r}_{ij})~ \hat{\mathbf{r}}_{ij} &;\hspace{0.1cm}\text{${r}_{ij}<\sigma$}\\
0 &;\hspace{0.1cm}\text{${r}_{ij}> \sigma$}.
\end{cases} 
\end{equation}
Here, $k_{eff}(ij)$ is the effective stiffness constant between $i^\text{th}$ and $j ^{th}$ particle and is defined as $ k_{eff}(ij) = \frac{k_i k_j}{(k_i + k_j)}$. Here, $i^{th}(j^{th})$ particle can be either hard or soft. When the $i^{th}$  particle is classified as \emph{hard}, its stiffness constant is assigned the value $k_i = k_h$. Conversely, if the particle is classified as \emph{soft}, its stiffness constant is set to $k_i = k_s$. The quantity $r_{ij}= |\boldsymbol{r}_i - \boldsymbol{r}_j|$, $\sigma$ is the effective diameter of the particle and $N = N_h+N_s$ is the total number of particles. 

We perform Brownian dynamics simulations with $N=3649$ to $233546$ particles. Taking $\sigma$ as the unit length, we define the  P\'eclet number (Pe) as the ratio of two length scales, $\text{Pe} = \frac{v_p}{\sigma~D_r}$. We fix the rotational diffusion coefficient $D_r = 0.005$, mobility $\mu=1$. Additionally, we define a non-dimensional stiffness constant, denoted as $\tilde{k} = \frac {\mu~ k}{D_r}$. In our simulations, the fraction of hard particles in the system is denoted $ x_h = \frac{N_h}{N_h+N_s}$ and the fraction of the soft particle as $x_s = 1 - x_h$. This system has been explored by changing the ratio of soft and hard particles while keeping the total number of particles the same. To quantify finite effects, we have also performed simulations for a different box  $L = \{ 128, 256, 512, 1024\}$. To see the effect of softness on overall collective behavior in the binary system, we ran simulations for $\tilde{k}_s=\{800, 1600, 2400, 4000\}$, keeping $\tilde{k}_h = 8000$. Since our focus is understanding the effect of particle composition in phase properties, we fix the total area fraction $\phi = \frac{(N_h+N_s)\pi~\sigma^2}{4L^2} =0.7$ in all our simulations.

To quantify the phase behaviour, we calculate the distribution of local number density of the particles $\phi_\text{loc}$. When the system phase separates, the $\phi_\text{loc}$ has a bi-modal distribution and the difference between peak positions of local density distribution is given by  $\Delta \phi = \phi_{loc}^{peak}(max) - \phi_{loc}^{peak}(min) $. If the system is homogeneous in density, then $\Delta \phi = 0$. The clustering in the system is quantified by calculating the largest cluster fraction, $f_L$, which is the fraction of the total number of particles in the largest cluster. A pair of particles form a cluster if the distance between them $|{\bf r}_{ij}|<\sigma$. We also quantify the structure of the dense phase by computing the global hexatic order which is defined as, $\psi_6 = \bigg\langle \left| \frac{1}{N} \sum \limits_{i=1}^{N} \phi_{6i} \right|\bigg \rangle,$ where $\phi_{6i}=\frac{1}{N_b} \sum \limits_{j \in N_b} e^{6 i \theta_{ij}},$ here $\theta_{ij}$ is the angle between ${\bf r}_{ij}$ and the reference axis, and $N_b$ represents the total number of neighbours of $i^{th}$ particle, calculated using Voronoi tessellation. 

\section{Results}
\subsection{Phase diagram}
    \begin{figure}[h!]
    \centering
    \includegraphics[width=0.48\textwidth]{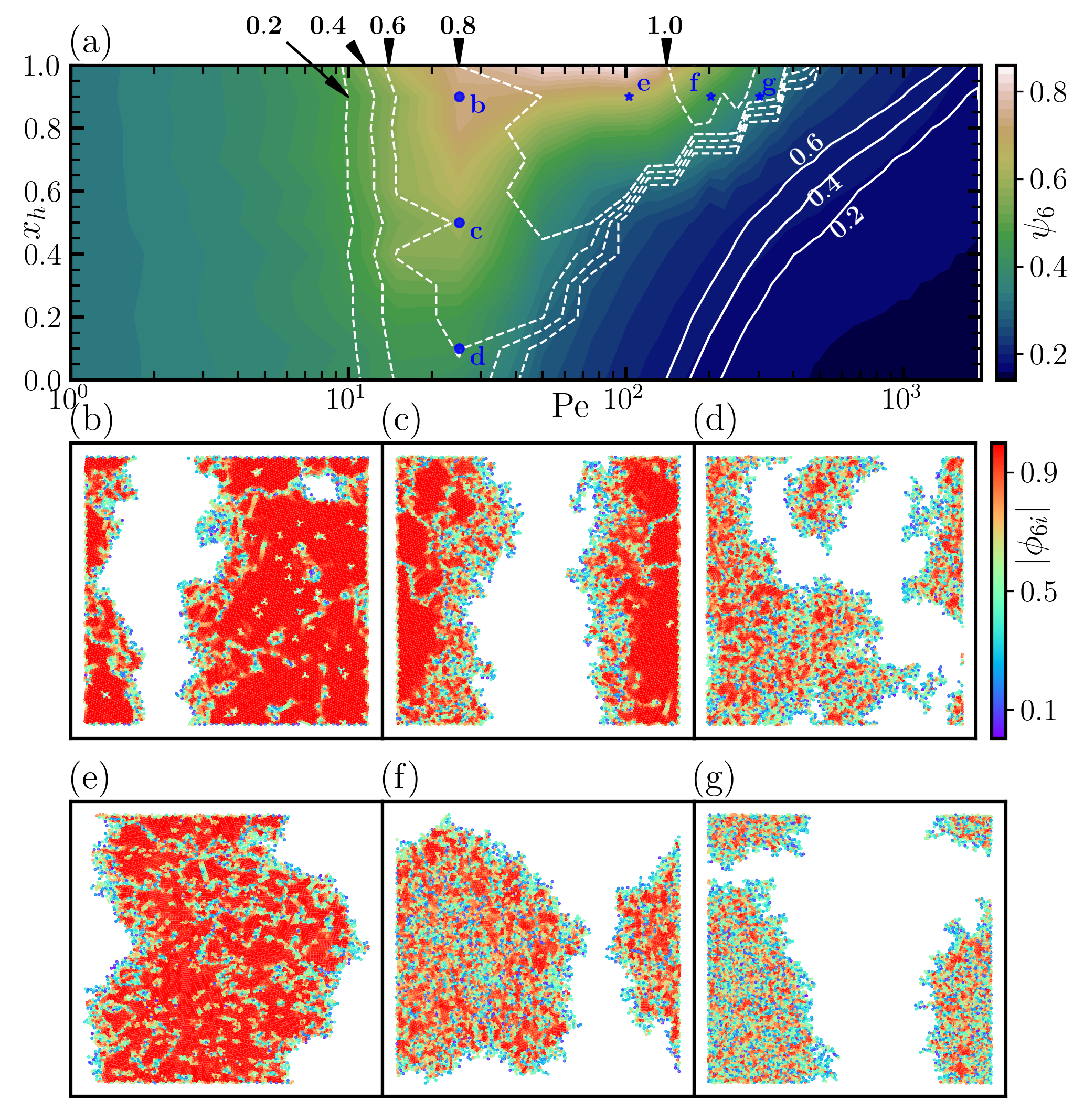}
    \caption{(a) Contour heat map of hexatic order in (Pe-$x_h$) plane for $\phi = 0.7$,  $N=10000$. The white solid lines indicate the order parameter cluster fraction $f_L$ for values $0.2,0.4, \text{and}~ 0.6$, along with that white dashed lines represent the contours of $\Delta \phi$ for $0.2,0.4,0.6,0.8, \text{and}~ 1.0$. 
    The configuration plot with steady-state plot, where the color code represents the local hexatic order $\phi_{6i}$ for (b-d) the three points \textbf{b}$\equiv(25,0.9)$, \textbf{c}$\equiv(25,0.5)$, and \textbf{d}$\equiv(25,0.1)$ with fix Pe$=25$ and \textbf{e}$\equiv(100,0.9)$, \textbf{f}$\equiv(200,0.9)$, and \textbf{g}$\equiv(300,0.9)$ (e-f) three with fix $x_h = 0.9$.} 
    \label{fig:4Fig1}
    \end{figure}
For particle stiffness $\tilde{k}_h = 8000$ and $\tilde{k}_s = 800$, we study the phase behaviour of ABPs as a function of Pe and the composition, $x_h$. In Fig~\ref{fig:4Fig1}(a), we plot the overall phase behaviour of the system in Pe- $x_h$ plane. The color map indicates the global hexatic order $\psi_6$. The solid white line represents the contours of $f_L= 0.2, 0.4, \text{and}~ 0.6$, indicating the parameter range in which $f_L$ is larger than the value indicated on the contour. Similarly, the white dashed lines represent the contours of $\Delta \phi = 0.2,0.4,0.6,0.8,1.0$; they demarcate the range of parameters in terms of the difference of the peak densities in this system. The MIPS state is characterized by a formation of dense clusters with a high hexatic order, indicated by a large value of $f_L$, $\psi_6$, and $\Delta \phi$. Indeed, a range of parameter values with $f_L > 0.6$ also provide large $\psi_6$ and $\Delta \phi$, indicating the presence of MIPS. However, the phase diagram also display two interesting features. 
First, for a broad range of parameters at large Pe, both $\Delta \phi$ and $\psi_6$ are small while $f_L >0.6$, indicating the formation of disordered clusters. 
Second, there exists a region for large Pe, marked by a large $\Delta \phi$ and $f_L$, but small $\psi_6$. Also, the regions of largest $\Delta \phi$ ($\Delta \phi > 1.0$, region $f$ in Fig~\ref{fig:4Fig1}(a)) does not coincide with maximum of the hexatic order $\psi_6 = 0.8$ for $x_h = 1.0$. A similar behaviour can be observed by reducing the value of $x_h$ for same Pe. While we explore the first feature in detail in the next section, we analyze the second feature by carefully examining the structural features representative configurations from different regions in the phase diagram where both $\Delta \phi$ and $f_L$ are large.

We examine a set of representative points: {\bf b}, {\bf c}, and {\bf d}, with corresponding values of $x_h = 0.9, 0.5, 0.1$ (or $x_s = 0.1, 0.5, 0.9$) respectively, while keeping Pe $= 25$ fixed. Additionally, we consider points {\bf e}, {\bf f}, and {\bf g} corresponding to Pe $= 100, 200, 300$, while keeping $x_h = 0.9$. These points clearly exhibit phase separation, as evidenced by $\Delta \phi > 0.2$. In Fig~\ref{fig:4Fig1}(b-g), the color-coded representation indicates the local order parameter $\phi_{6i}$ for each of these points.


The first three points correspond to a fixed Pe for different $x_h$, which are explained in the details: (i) For large proportion of hard particles $x_h=0.9$ (point {\bf b}), Fig~\ref{fig:4Fig1}(b) shows the formation of the single large cluster and high local hexatic order. (ii) For an equal number of hard and soft particles $x_h=x_s=0.5$ at point {\bf c}, Fig~\ref{fig:4Fig1}(c) shows the formation of the MIPS. However, the size of the MIPS cluster is reduced, and also, one sees a decrease in the local order. (iii) For mostly soft particles in the system $x_h=0.1$, point {\bf d}, Fig~\ref{fig:4Fig1}(d) shows a further decrease in cluster and decrease in local order. 
This shows that with the increase in softness and by decreasing $x_h$, both the MIPS cluster and the local hexatic order decrease.
To observe the impact of Pe, we considered additional points {\bf e}, {\bf f} and {\bf g} in addition to point {\bf b}, while maintaining a constant particle fraction of $x_h = 0.9$ and varying Pe values. (iv) At point {\bf b} in Fig~\ref{fig:4Fig1}(b) where Pe $= 25$, we observed the formation of a large MIPS cluster with a high hexatic order. However, the range of $\Delta \phi$ remained moderate, between $0.6$ and $0.8$. (v) Fig~\ref{fig:4Fig1}(e) represents point {\bf e}, corresponding to Pe value, where we observed a reduction in the size of the largest MIPS cluster as well as a decrease in its local hexatic order. However, we noticed an increase in the range of $\Delta \phi$, between $0.8$ and $1.0$. (vi) For point {\bf f}, despite having a smaller MIPS cluster size and lower local hexatic order, $\Delta \phi$ exceeded $1.0$, reaching its highest value. (vii) Point {\bf g} lies within the range of $1.0 > \Delta \phi > 0.8$, indicating a further decrease in the size of the MIPS cluster accompanied by a decrease in local order.
This demonstrates that as Pe increases, indicating increased softness, the behavior of $\Delta \phi$ is nonmonotonous. Both the cluster size and local hexatic order decrease, and the parameters corresponding to high hexatic order do not  coincide with the largest $\Delta \phi$.
From the above analysis, it is evident that the phase-separated dense clusters can have a disordered structure for certain parameter regions, leading to a low $\psi_6$ albeit $f_L$ and $\Delta \phi$ are large.

\subsection{Effect of particle softness}
    \begin{figure}
    \centering
    \includegraphics[width=0.48\textwidth]{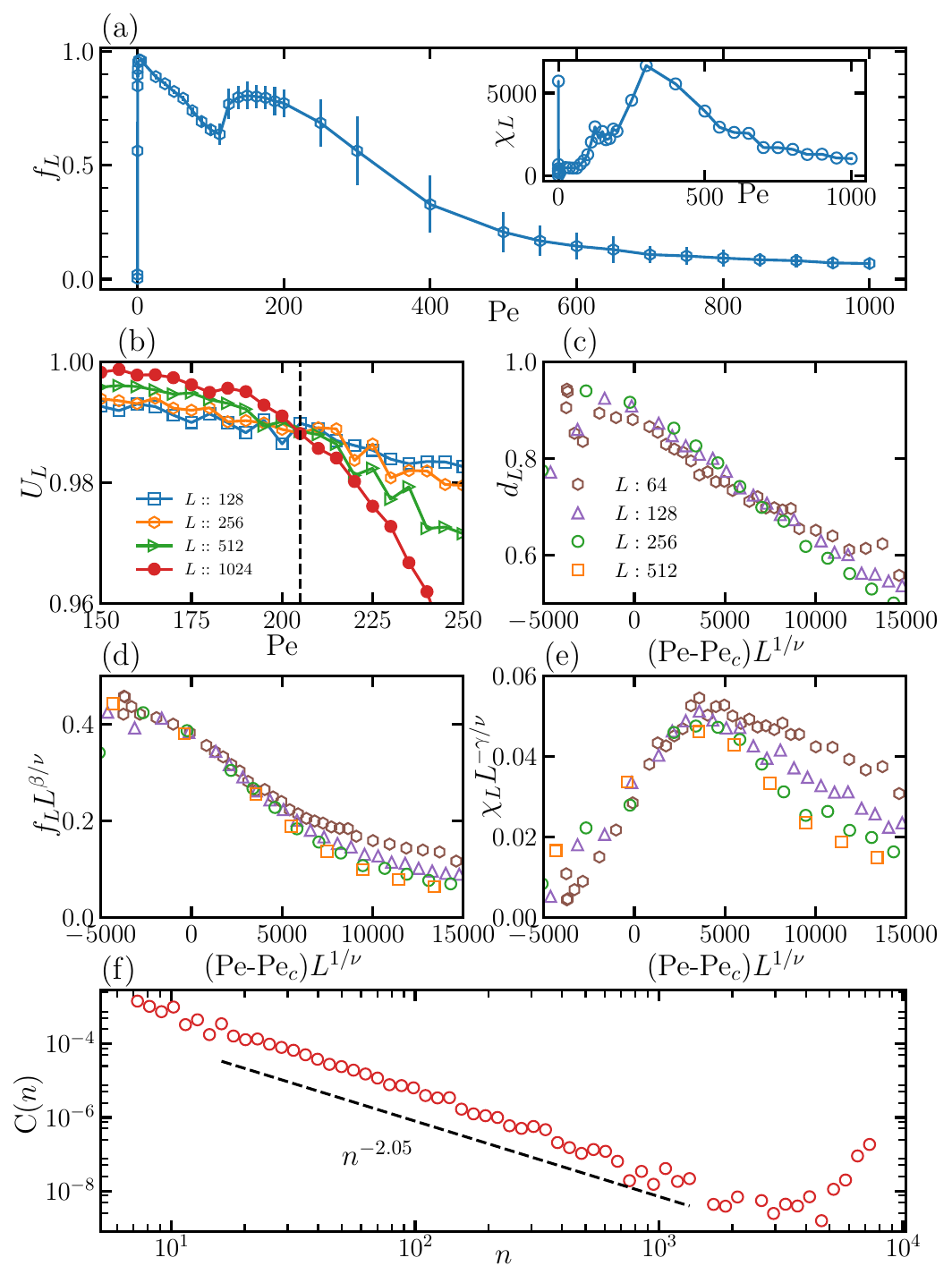}
    \caption{Percolation Transition: (a) the mean largest cluster fraction $f_L$ and susceptibility  $\chi_L$ in the inset. (b) Fourth order binder cumulant $U_L$, finite scaling of order parameter (c)average linear extension $d_L$, (d) largest cluster fraction $f_L$, (e) susceptibility $\chi_L$ with system  $L = 128, 256, 512, 1024$. This collapse occurs at the exponents values $\nu \approx 1.42$, $\beta \approx 0.14$, and $\gamma \approx 2.8$. $\text{Pe}_c \approx 205$.(f) cluster  distribution at $\text{Pe}_c \approx 200$ which scale with $2.1$.}
    \label{fig:4Fig2}
    \end{figure}
From the phase diagram in Fig~\ref{fig:4Fig1}(a), it is evident that for a fixed value of $x_h$, the overall phase behavior changes with an increase in Pe.
To systematically study these changes, we fix $x_h= 0.5$ and plot the corresponding $f_L$ as a function of Pe (see Fig~\ref{fig:4Fig2}(a)). Similar to homogeneous system results (see Fig~3 in \cite{sanoria2022Percolation}), the order parameter $f_L$ shows non-monotonous behavior and exhibits multiple phase transitions. This is revealed by the peaks in the susceptibility $\chi_L = L^2~\sqrt{\langle f_L ^2 \rangle - \langle f_L \rangle ^2}$, indicating the presence of three transitions (see inset Fig~\ref{fig:4Fig2}(a)). For the homogeneous system, the last transition at large Pe was demonstrated to be a percolation transition~\cite{sanoria2022Percolation}. We closely analyze the large Pe region to study how  the transition in this region is influenced by the heterogeneity in the system.
For a systematic characterization of this transition, we performed a finite size  analysis with system sizes $L = 128, 256, 512$, and $1024$. 
Near the critical point, the order parameter $f_L$, and susceptibility $\chi_L$ should follow the scaling relations $f_L=L^{-\beta/\nu}f((\mbox{Pe}-\mbox{Pe}_c)L^{1/\nu})$, $\chi_L=L^{-\gamma/\nu}g((\mbox{Pe}-\mbox{Pe}_c)L^{1/\nu})$, respectively. Here, $f,~g$ are scaling functions, and $\beta, \nu,$ and $\gamma$  are the universal exponents.
We first estimate the critical parameter Pe$_c$ by calculating the fourth order Binder cumulant {$U_L=\frac{1}{2}~(3 - \frac{\langle d_L^4 \rangle}{\langle d_L^2 \rangle^2})$}, where $d_L$ is the normalized maximum extension of the largest cluster, defined as $\frac{\langle l_m \rangle}{\sqrt{2}L}$, where $l_m$ is twice the distance from the centre-of-mass of the largest cluster to the particle at the farthest point within the cluster and $L$ is the box size~\cite{kyriakopoulos2019clustering}. In Fig.~\ref{fig:4Fig2}(b), we show the binder cumulant $U_L$ as a function of Pe which gives us an estimate of Pe$_c \approx 205$.
We find that for Pe$_c=205$ both $f_L$ and $\chi$ collapses for values of $\nu \approx 1.42,~\beta \approx 0.14,~\mbox{and}~\gamma \approx 2.8 $ (shown in Fig~\ref{fig:4Fig2}(c-e)).
These exponent values are in reasonable agreement with the universal exponents for standard 2D percolation, which are $ \nu = 1.33,~\beta = 0.14,~\mbox{and}~\gamma = 2.39.$  This agreement in critical exponents obtained for the estimated Pe$_c$ is a strong indicator of a percolation transition, similar to that of a homogeneous system \cite{sanoria2022Percolation}. We also plot the cluster  distribution $c(n)$ at a critical point we see that $c(n)$ scales approximately as $n^{-2.05}$ (Fig~\ref{fig:4Fig2}(f)). Again, the exponent value is in good agreement with the Fisher exponent for the standard percolation transition $2.05$. This establishes that an active binary mixture with an equal proportion of soft particles with different stiffness also exhibits a percolation transition at the high motility similar to the homogeneous case. 

\begin{figure}[h!]
\centering
\includegraphics[width=0.48\textwidth]{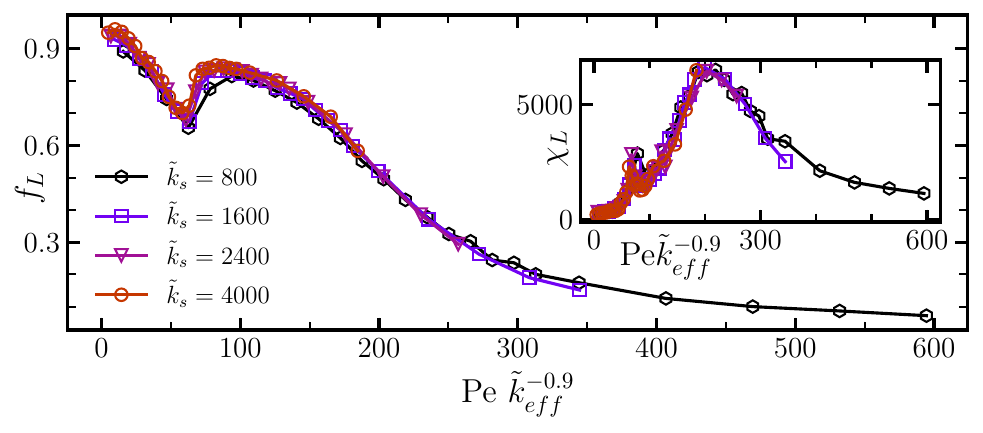}
\caption{Overall cluster fraction $f_L$ plots for different $
\tilde{k}_s$ values scales as $\tilde{k}_{eff}^{-0.9}$, its susceptibility in the inset also scaled with $\tilde{k}_{eff}^{-0.9}$. $\tilde{k}_{eff}$ is the effective interaction strength between the dissimilar types of particles, calculated using $\tilde{k}_h$ and $\tilde{k}_s$. } 
\label{fig:4Fig3}
\end{figure}
We further study the phase properties of the system by varying the interaction strength of the soft particles ($\tilde{k}_s$) for a fixed composition, $x_h=0.5$ and $\tilde{k}_h = 8000$. In Fig.~\ref{fig:4Fig3}
we plot $f_L$ and $\chi_L$ as a function of Pe, for different values of  $\tilde{k}_s$. We find that with the increase in $\tilde{k}_s$, the transition point of various phases shifts to the higher value of the Pe. However, upon scaling the abscissa (Pe) with $ \tilde{k}_{eff}^{-0.9}$, we obtain a very good data collapse.
The same scaling is also followed by the susceptibility  $\chi_L$ (see inset of Fig~\ref{fig:4Fig3}). We note that the exponent obtained in this case ($\sim \tilde{k}_{eff}^{-0.9}$) is different from the scaling behaviour obtained in homogeneous ABPs \cite{sanoria2022Percolation}, where an exact inverse relationship between Pe and $k$ was obtained.
This difference in the exponent is clearly due to the heterogeneity in the system.

\subsection{Effect of variation in composition}
    
\begin{figure}[h!]
    \centering
    \includegraphics[width=0.48\textwidth]{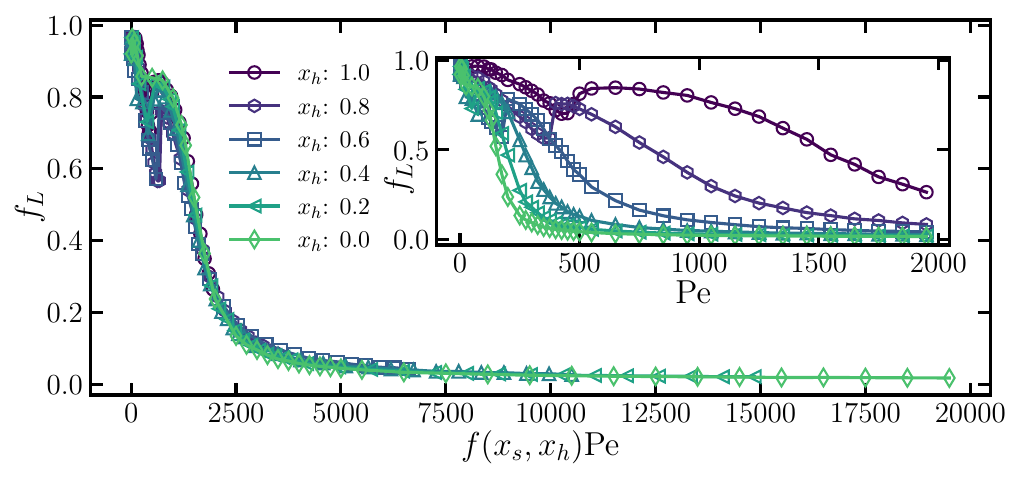}
    \caption{The mean largest cluster fraction $f_L$ as a function of Pe for various values of $x_h = 1$ to $0$ (inset) scaled with non-linear function $f(x_s,x_h) = {k_h \over k_s}x_s^\alpha + x_h ^\beta$ where $\alpha \approx 1.22$ and $\beta \approx 4.16$.}
    \label{fig:4}
    \end{figure}

    In the previous section, we have shown that the softness dependent overall phase behavior is qualitatively similar to the homogeneous system, while the scaling with the effective stiffness is marginally different. Changing the composition of the system ($x_h$) while keeping all other parameters intact can also cause changes in the overall mechanical properties of the system. To study the effect of composition in the overall phase behaviour, we vary $x_h$ from $0$ to $1$, with $\tilde{k}_h=8000$ and $\tilde{k}_s=800$. When $x_h =1$ the system contains a collection of stiff particles ($\tilde{k} = 8000$) and we observe the percolation transition at a high (Pe$_c \approx 550$) value. However, as we decrease the $x_h$, i.e., upon partially replacing hard particles with soft particles, the transition points gradually shift to the lower value of Pe (Pe$_c \approx 205$ for $x_h=x_s=0.5$) (inset of Fig~{\ref{fig:4}}). When $x_h =0$ the system again becomes homogeneous, consisting of only soft particles of $\tilde{k}=800$. 
We found that the overall behavior of $f_L$ in Fig.~\ref{fig:4} is nicely scaled with a non-linear function $f(x_s,x_h) = {k_h \over k_s}x_s^\alpha + x_h ^\beta$, where $\alpha \approx 1.22$ and $\beta \approx 4.16$. This indicates that the unequal proportion of hard and soft particles significantly affects the interparticle interaction; a simple, effective spring constant cannot replace them.
 
\subsection{Structure and composition of clusters}
    \begin{figure}[h!]
    \centering
    \includegraphics[width=0.48\textwidth]{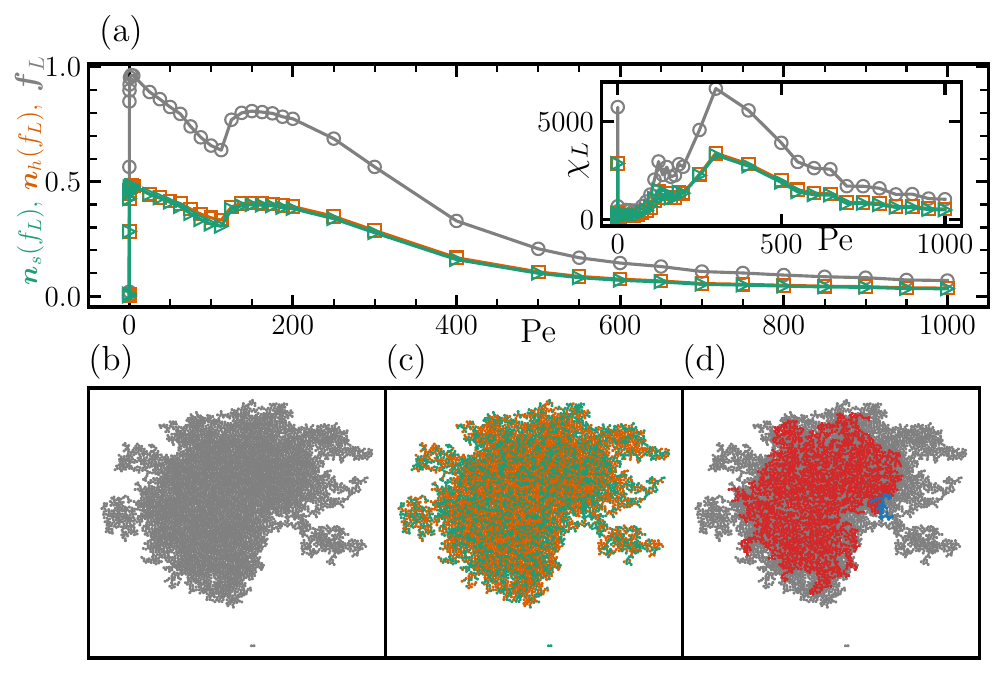}
        \caption{Study of largest cluster composition.
        (a) Largest cluster fraction $f_L$, total number fraction of hard particles $n_h(f_L)$, and of soft particles  $n_s(f_L)$ plotted as a function of Pe, corresponding susceptibility plots in inset of (a). 
        Overall largest cluster (a) without differentiating between the type of particles, (b) hard and soft particles marked with color orange and green, respectively. In (c) we show the largest cluster within dense cluster for hard and soft particles separately with red and blue colors, respectively at Pe=125. These simulations are performed at constant $\tilde{k}_h = 8000$ and $\tilde{k}_s = 800$}
    \label{fig:4Fig5}
    \end{figure}

 After studying the overall phase behaviour of the system, we now study the composition and distribution of soft and hard particles within the particle clusters. To quantify this, we first separately calculate $n_h(f_L)$ and $n_s(f_L)$ which represents the fraction of hard and soft particles in the largest cluster fraction $f_L$. We consider the case for $x_h =0.5$, with an equal proportion of hard and soft particles, for fixed $\tilde{k}_h = 8000$, and $\tilde{k}_s = 800$. Fig~\ref{fig:4Fig5}(a) we plot $n_h(f_L)$ and $n_s(f_L)$ as a function of Pe and compare it with $f_L$, the overall largest cluster fraction. It is evident from the figure that both $n_h(f_L)$ and $n_s(f_L)$ are almost equal for all Pe, roughly $f_L/2$. Apparently, an equal number of particles in the system leads to an equal proportion of particles in the clusters. A similar behaviour is also observed when the susceptibility $\chi_n^{h}$ and $\chi_n^s$ are calculated (Fig~\ref{fig:4Fig5}(a)-inset). However, the distribution of both types of particles inside a dense cluster shows significant differences. In Fig.~\ref{fig:4Fig5}(b)-(d) we study this particle distribution for Pe=125, where the system is in the MIPS state. Fig~\ref{fig:4Fig5}(b) shows the largest cluster formed at Pe$=125$, which includes both hard and soft particles. In Fig~\ref{fig:4Fig5}(c) we separately label both types of particles inside the dense cluster. The particles are clearly intermixed within the dense cluster, showing no sign of stiffness-induced segregation. In Fig~\ref{fig:4Fig5}(d) we indicate the largest cluster, made of both types of particles separately, along with the whole cluster. While the largest cluster made of hard particles is comparable in size with the whole cluster, the largest cluster of soft particles is much smaller in size. Interestingly, the hard particles separately form a large, porous network inside the dense cluster and the soft particles form small disconnected clusters, while the overall particle density of such a cluster is quite large. This behaviour is observed for a range of Pe, as described below. 

\begin{figure}[h!]
    \centering
    \includegraphics[width=0.48\textwidth]{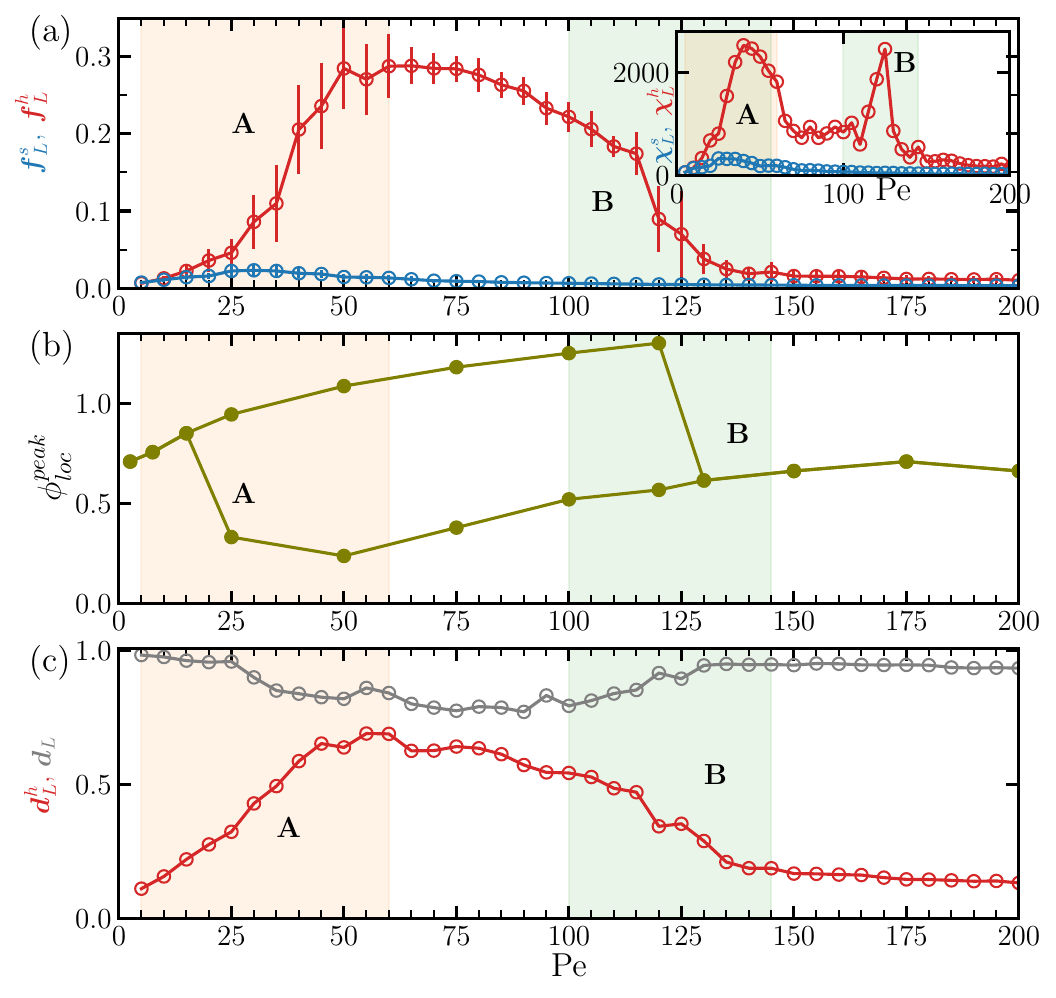}
        \caption{ (a) Fraction of the largest cluster of hard ($f_L(x_h)$) and soft particles ($f_L(x_s)$) as function of Pe, color scheme followed of Fig.~\ref{fig:4Fig5}(c),and the susceptibility $\chi_L^h$ of $f_L(x_h)$, and $\chi_L^s$ of $f_L(x_s)$ is plotted as a function of Pe (see inset). (b) The density corresponding to peak value of the local density distribution $\phi_{loc}^{peak}$ is plotted as a function of Pe for $x_h= x_s = 0.5$. (c) Normalised maximum extension $d_L$ for overall largest cluster $f_L$ and normalised maximum linear extension $d_L(x_h)$ for $f_L(x_h)$ plotted for Pe. These simulations are performed at constant $\tilde{k}_h = 8000$ and $\tilde{k}_s = 800$} 
    \label{fig:4Fig6}
    \end{figure}

The formation of a separate space-filling cluster by the hard particles implies that they form a porous scaffold inside the dense cluster and the soft particles get filled in by forming smaller clusters. 
To explore this behaviour quantitatively, we separately calculate the largest cluster fraction of hard ($f_L^{h}$) soft particles ($f_L^{s}$) as a function of Pe. 
    As shown in Fig~\ref{fig:4Fig6}(a), the value of $f_L^{s}$ remain small for all the Pe. Interestingly, $f_L^{h}$ shows non-monotonous behavior as its value steeply increases in region-A and decreases in region-B (Fig~\ref{fig:4Fig6}(a)), indicating that the hard particles form large clusters in that parameter range.  
     We also separately calculate the corresponding susceptibilities $\chi_L^{h}$ and $\chi_L^{s}$ using $f_L^h$ and $f_L^{s}$, respectively (see inset of Fig~\ref{fig:4Fig6}(a)). Two distinct peaks of $\chi_L^{h}$ in region-A and region B and an absence of any clear peaks in $\chi_L^{s}$ again indicate the presence of transitions involving only hard particles. 
     To obtain a complete picture of the hard particle arrangement in the context of overall phase behaviour of the system, we compare local density corresponding the peak values of the distribution, $\phi_{loc}^{peak}$ in Fig~\ref{fig:4Fig6}(b).  As shown in \cite{sanoria2022Percolation}$, \phi_{loc}^{peak}$ is single-valued only  when the system is homogeneous and it bifurcates when the system is in the MIPS state.
    By comparing Fig~\ref{fig:4Fig6}(a) and Fig~\ref{fig:4Fig6}(b) we see that region with high $f_L^{h}$ ($15 \lesssim$Pe $ \lesssim 130$) coincides with region Pe where density bifurcation takes place in $\phi_{loc}^{peak}$. The presence of a MIPS state implies the formation of a large, compact cluster consist of both hard and soft particles in equal proportion, whereas a large $f_L^{h}$ indicates the presence of a connected cluster of hard particles. The simultaneous presence of both these entities is a strong indication of a porous hard-particle cluster {\it inside} the dense MIPS cluster.  

We further examine the structural arrangement of particles in this parameter range,  by calculating the normalized, maximum linear extension $d_L^{h}$ for the largest cluster of the hard particles $f_L^{h}$. As evident in Fig.~\ref{fig:4Fig6}(c), $d_L$ starts to decrease in region-A and reaches a minimum at $50<$Pe$<100$. In the same region $d_L^h$ increases  and reaches a maximum. A decrease in $d_L$ indicates that the overall cluster becomes more compact, whereas the increase in $d_L^h$ shows a spatial growth of the hard particle cluster. The minimum of $d_L$ coincides with a maximum in $d_L^h$, and the values of both these entities become comparable at this Pe range. Since the composition of both types of particles is equal inside the cluster, this observation verifies the formation of a space-filling cluster of hard particles  and several small clusters of soft particles inside the overall dense cluster, for a range of Pe.

\section{Summary}

This work demonstrates that introducing heterogeneity via inter-particle interactions leads to a more complex overall phase behaviour for active Brownian particles. Especially at large Pe, the system also phase separates to form dense, stable clusters with low hexatic order in addition to forming ordered dense clusters at slightly lower Pe. The presence of such clusters indicates that introducing soft particles destroy the crystalline order, without affecting the stability of the cluster. However, a further increase in Pe destabilises the dense clusters and subsequently forms a porous cluster, similar to the homogenous systems~\cite{sanoria2022Percolation}. We show that the formation of the porous cluster in this case is also associated with a percolation transition, as observed for homogeneous systems. 

Similar to our previous work \cite{sanoria2022Percolation}, here we have systematically characterized the percolation transition at high Pe by calculating the critical parameter Pe$_c$ of the transition and the critical exponents using a finite-size scaling analysis. The exponents estimated by collapsing the order parameters and the susceptibility are in good agreement with the critical exponents of the 2D percolation transition.

While the overall phase behaviour, including phase separation and the percolation transition, is qualitatively similar to the homogenous case, new interesting macroscopic properties were observed due to an interplay between particle stiffness, composition, and activity. Keeping the composition equal, a decrease in stiffness of the soft particles ($k_s$) shifts the overall phase behaviour to a lower range of Pe. We show that this shift scales almost linearly with an effective particle stiffness ($\sim k_{eff}^{-0.9}$). A similar shift in phase behaviour to a lower range of Pe is also observed when the proportion of soft particles ($x_s$) is increased. We show that this shift can be scaled using a non-linear function $\sim {k_h \over k_s}x_s^\alpha + x_h^\beta$. The absence of linear scaling, in spite of linear elastic interparticle interactions, indicates the emergence of complex macroscopic mechanical properties due to heterogeneity at the microscopic level.

A detailed analysis of the composition and the structure of dense clusters further revealed several novel features. We found that the proportion of soft/hard particles of the dense cluster is the same as the overall particle composition in the whole system. However, inside the dense cluster, the hard particles form a spanning porous network while the soft particles fill the pores by forming smaller clusters. This formation of a porous network of hard particles inside a dense cluster is reminiscent of a rigid skeleton, providing stability to the structure and supporting the soft particles filling the pores. 
In summary, we have shown that for an ABP system, the heterogeneity in composition can significantly affect the phase behavior. 
\emph{Acknowledgement:}~ RC acknowledges the financial support by Science and Engineering Research Board (SERB), India via Project No. CRG/2021/002734. We thank IIT Bombay HPC facility (Spacetime2).

\bibliography{ref}
\end{document}